\documentclass[preprint]{elsarticle}
\usepackage[utf8]{inputenc}
\usepackage{amssymb}
\usepackage{graphicx} 
\usepackage{lineno}
\usepackage[hidelinks]{hyperref}
\usepackage{natbib}

\journal{ArXiv}

\begin{document}
\begin{frontmatter}


\title{On the prospects of enhance glass based devices with silver nanoparticles: the case of $Pr^{3+}$ doped tellurite-tungstate glasses}


\author{Maiara Mitiko Taniguchi\corref{cor1}$^a$}
\ead{maaymt@gmail.com}
\cortext[cor1]{Corresponding author}	
\author{Jorge Diego Marconi$^b$}
\author{Leandro Silva Herculano$^c$} 
\author{Marcos Paulo Belan\c{c}on$^a$} 

 \address[$^a$]{P\'{o}s-graduação em Processos Qu\'{i}micos e Bioqu\'{i}micos, Universidade Tecnol\'{o}gica Federal do Paran\'{a}, Via do Conhecimento Km 1, 85503-390 Pato Branco - Brazil}
\address[$^b$]{Centro de Engenharia, Modelagem e Ciências Sociais Aplicadas, Universidade Federal do ABC, Santo Andr\'{e}, SP, Brazil}
\address[$^c$]{Departamento de F\'isica, Universidade Tecnol\'{o}gica Federal do Paran\'{a}, Medianeira, PR, Brazil}

\begin{abstract}


The development of enhanced optical devices relies on key materials properties. In this context, researchers have investigated the effects of metal nanoparticles incorporation in active rare-earth materials. Here, we discuss the effects of Silver nitrate addition 0.5\%, 1.0\% and 2.0\% mol\%) into a $Pr^{3+}$ doped tellurite-tungstate glasses (TWNN) formed by the conventional melt-quenching technique. The XRD curves confirmed the amorphous structural nature of TWNN glasses. The FTIR spectra identified the presence of different vibrational groups. DSC curves verified the increase around 70~$^{\circ}$C for thermal stability with silver addition. Absorption spectra observed surface plasmon resonance (SPR) band around 500 nm. Luminescence measures shows enhanced and quenching for $Pr^{3+}$ transitions under the 473 nm excitation. The band around 480-520 nm are more affected by SPR band by quenching, and the bands around 575-675 nm are luminescence emission increased due to the presence of silver nanoparticles. In addition, lifetime decays as a function of heat treatment time, has also been verified. The Ag enhances the lifetime regarding the samples annealed by 20 hours and as the Ag concentration increases, the lifetime stretches as well. The present results shows that $Pr^{3+}$-Ag codoped TWNN glass lead to some remarkable improvement, verified in properties such as lifetime constants or emissions intensities and is a promising candidate for photonic devices.
\end{abstract}


\end{frontmatter}


\section{Introduction}
\label{sec:intro}

In a previous work we have presented some interesting properties in Pr$^{3+}$ doped tellurite-tungstate (TWNN) glasses, which could have its near-infrared emission tuned when the glass is codoped with Yb$^{3+}$\cite{Belancon2014a}. Aiming further improvements in those optical properties we have investigated possible enhancements due the presence of silver nanoparticles (AgNPs) around the rare-earth (RE), and this work presents some discussion about the results and challenges we have observed.

The rich energy diagram of Pr$^{3+}$ provides a huge potential for several applications. In the other hand,  the high number of metastable states makes possible to depopulate upper levels by many different pathways that are very difficult to avoid. In some RE ions, like Er$^{3+}$ for instance, it is possible to improve intensities selectively by co-doping the glass with a proper ion that quench some specific metastable state. We can then enhance the probability of one transition by reducing another. However, this approach is not easily applied to Pr$^{3+}$ due the overlap between multiple transitions. 

For example, in our TWNN glass\cite{Belancon2014a} at least three luminescence bands are originated from the $^1D_2$ level, at the wavelengths of $\sim$600 nm, $\sim$1050 nm and $\sim$1480 nm\cite{Belancon2014a}. This last one matches the ground state absorption to the $[^3F_3,^3F_4]$ manifold, which is pretty wide and overlaps the emission at $\sim$1300 nm from the $^1G_4$ manifold\cite{Ohishi1991}. As the upper excited states of Pr$^{3+}$ ($^{3}$P$_{2}$, [$^{3}$P$_{1}$,$^{1}$I$_{6}$], $^{3}$P$_{0}$) have about twice the energy of the $^1G_4$, multiple cross-relaxation process may take place.

To avoid this self-quenching, the Pr$^{3+}$ concentration should be kept at quite low levels, even in comparison with other rare-earth ions incorporated in the same glass. For example, optical amplification in Er-Yb codoped TWNN glass was already demonstrated\cite{Narro-Garcia2013} for a total dopant concentration reaching 3\%. Meanwhile, our previous study on Pr$^{3+}$ indicates that self-quenching is already effective at concentration as low as 0.1\% in single doped samples. 

Aiming to enhance gain properties, researchers have broadly investigated the effects of co-doping active materials with silver nanoparticles. Furthermore, improvements in optical transitions have been verified in Er$^{3+}$\cite{Fares2014,Amjad2015,Rivera2010}, Nd$^{3+}$\cite{Yraola2015}, Eu$^{3+}$\cite{Raja2014}, Dy$^{3+}$\cite{RezaDousti2014}, Sm$^{3+}$\cite{Yusoff2015} and Pr$^{3+}$\cite{Lakshminarayana2009,kassab2007,Du2015203}, among others, where authors have frequently interpreted that in terms of enhanced electric field around the rare-earth ions due the presence of nanoparticles (NPs). 

Although this phenomena may favors the intensity of some transitions, practical demonstrations of optical devices exploiting REs co-doped with AgNPs are until now restricted to sensing surfaces\cite{Raj2016, Ortega-mendoza2014} or special solid states laser designs\cite{Molina2016}, which are not based in the traditional gain media that can be melted, quenched and fiberized. Moreover, it was not found glasses codoped with RE and AgNPs applied in commercial lasers, amplifiers or optical converters in the literature.

The TWNN glass can be fiberized, and amplification in a RE doped TWNN fiber was already demonstrated\cite{Chillcce2006a}. By this way, given the context presented here, we performed this study to verify if we can improve optical properties of Pr$^{3+}$ doped TWNN glasses by introducing silver NPs without undesirable effects in the fiberization ability and in the performance of a practical device.

\section{Experimental procedure}
\label{sec:method}

The TWNN glass system is composed by $72.5{TeO_2}-23{WO_3}-3{Na_2CO_3}-1.5{Nb_2}{O_5}$ in mol\%. The glass was doped with 0.1${Pr_{6}}{O_{11}}$ and x${AgNO_3}$, where x=0.5, 1.0 and 2.0 mol\%. The powder materials with high purity ($>$99.9\%), were weighed ($\pm0.1$ mg) in appropriate proportions in order to obtain a total mass of 10 g and then manually mixed. The powders were preheated in a platinum crucible at 200 $^{\circ}$C for an hour to remove water residues and then melted at 700~$^{\circ}$C by an hour in open atmosphere, followed by quenching to around 300~$^{\circ}$C, just below the glass transition temperature (around 350~$^{\circ}$C). At this stage the sample were poured in different steel molds, and they were annealed at 300~$^{\circ}$C by periods of time ranging between a few minutes and 20 hours before cooling to room temperature.

X-ray diffraction (XRD) were obtained from the glass powder using a Rigaku Miniflex 600 diffractometer with CuK${\alpha}$ radiation of wavelength 1.5418~${\textup{\AA}}$ with 30 kV and 15 mA current.\ The scan range was set to 2$\theta$ from 5$^{\circ}$ to 80~$^{\circ}$ with a step size of 0.02$^{\circ}$.\ The differential scanning calorimetry (DSC) data was acquired with a TA Instruments DSC Q20 using N${_2}$ atmosphere, with flow rate of 50 mL min$^{-1}$ and heating rate of 10$^{\circ}$C min$^{-1}$. A platinum crucible was used and the samples were heated until 650 $^{\circ}$C. Infrared transmittance spectra were obtained using Fourier transform infrared (FT-IR) spectroscopy in attenuated total reflectance (ATR) mode in a ``Perkin Elmer Frontier''. The spectra were collected in the range 4000-400 cm$^{-1}$ with a resolution of 2 cm$^{-1}$ and averaged 32 scans. Visible and near infrared photoluminescence (PL) spectra were obtained with a Thorlabs CCS100 spectrometer under diode laser excitation. Luminescence decays were collected using a Jobin–Yvon Fluorolog3 spectrofluorimeter with a Xenon lamp (power of 450 W) or a nanoled as the pump source.



	


\section{Results and discussion}
\label{sec:results}

\subsection{Structural analysis}

\begin{figure}[h]
\centering
\includegraphics[scale=0.5]{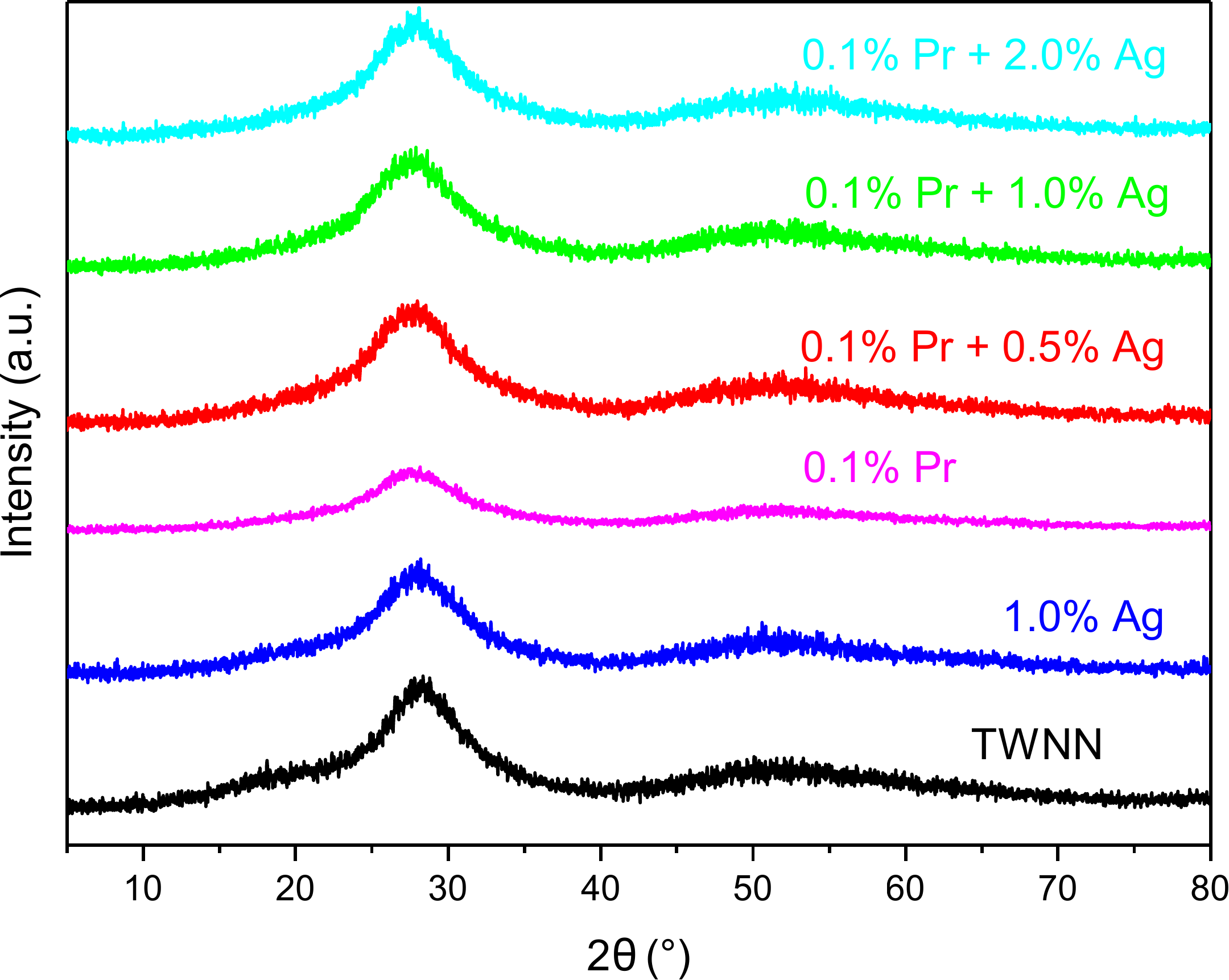}%
\caption{X-ray diffraction (XRD) profiles of TWNN samples annealed by 20 hours.}
\label{fig:XRD}
\end{figure}

Figure~\ref{fig:XRD} shows the XRD profiles for the TWNN glass and doped samples, all annealed by 20 hours. All samples exhibit similar XRD patterns, evidencing that the dopants have not induced structural modifications\cite{RAJESH2017607, WU2016185}. No sharp peak was observed, which indicates the absence of periodicity of the three-dimensional network in long range, typical of crystalline materials \cite{Soltani2016, WU2016185, Jayasimhadri2008}. It was observed  only a broad peak around 27~$^{\circ}$ of 2$\theta$, which is characteristic of the amorphous nature of the tellurite glasses \cite{Fares2014, Hssen2015, CHENG2017102}.

\subsection{Thermal analysis}

\begin{figure}[!htb]
\centering
\includegraphics[scale=0.5]{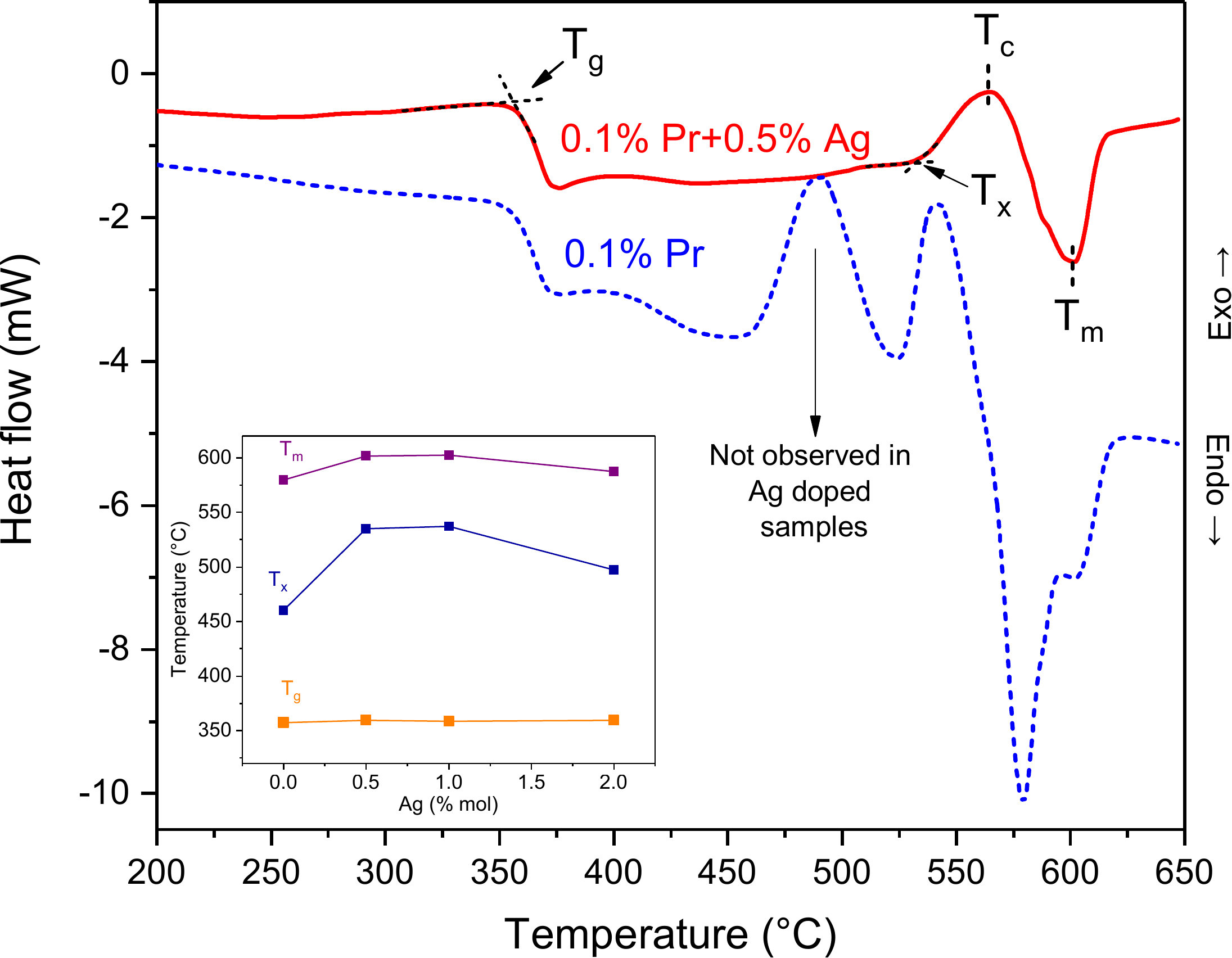}%
\caption{DSC curve of TWNN samples with 0.5mol\% of Ag and the inset shows the $T_g$, $T_x$, $\Delta$T and $T_m$ of as function of Ag concentration annealed by 20 hours.}
\label{fig:DSC_Ag}
\end{figure}

In Figure~\ref{fig:DSC_Ag} we can see DSC curves for 0.1\% $Pr^{3+}$ single doped and 0.5\% Ag codoped samples.\ The crystallization around 480~$^{\circ}$C is observed only in Pr$^{3+}$ singled doped sample, by this way all Ag doped samples have significant higher temperature of onset ($T_x$) and peak ($T_c$) of crystallization. In the inset we can see the critical temperatures as function of Ag concentration, and in Table \ref{tab:dsc} we have there values as well the thermal stability ($\Delta T=T_x-T_g$).

\begin{table}[!htb]
\centering
\caption{DSC parameters of all TWNN glasses with and without Ag.}
\label{tab:dsc}
\begin{tabular}{l c c c c}
\hline
\textbf{Sample} & \textbf{T$_g$ ($^\circ$C)} & \textbf{T$_{x}$ ($^{\circ}$C)} & \textbf{T$_{m}$ ($^{\circ}$C)} &  \textbf{$\Delta$T=T$_x$-T$_g$ ($^{\circ}$C)} \\
\hline

TWNN 0.1\% Pr & 357.31 & 460.53 & 579.59 & 103.22 \\
TWNN 0.1\% Pr 0.5\% Ag & 359.29 & 534.84 & 601.68 & 175.55 \\
TWNN 0.1\% Pr 1.0\% Ag & 358.77 & 537.27 & 602.57 & 178.50 \\
TWNN 0.1\% Pr 2.0\% Ag & 359.47 & 497.33 & 587.19 & 137.87 \\

\hline
\end{tabular}
\end{table}

As one can see, due the suppression of the exothermic event around 480~$^{\circ}$C in Ag doped samples, their thermal stability has improved. This is an important parameter of the potential to manufacturing devices, namely of optical fibers, where the glass will undergo a reheating process.\ Being subjected to heating cycles, undesirable crystallization may occur and even in small quantities the crystallization may cause significant light scattering, \textit{i.e.}, attenuating the optical signal and compromising the gain \cite{CHENG2017102,paz2016} achieved in amplifiers. 

In general, a glass with $\Delta$T greater than 100~$^{\circ}$C can be used as a feedstock for the production of optical fibers \cite{Soltani2016,DWIVEDI2015202}.\ As can be seen from Table~\ref{tab:dsc} the addition of Ag increased the stability in about 70~$^{\circ}$C. Such effect is significantly higher than the observed in other similar glasses\cite{CHENG2017102}.

\subsection{Optical analysis}

\subsubsection{Fourier Transform Infrared analysis}

\begin{figure}[htb]
\centering
\includegraphics[scale=0.5]{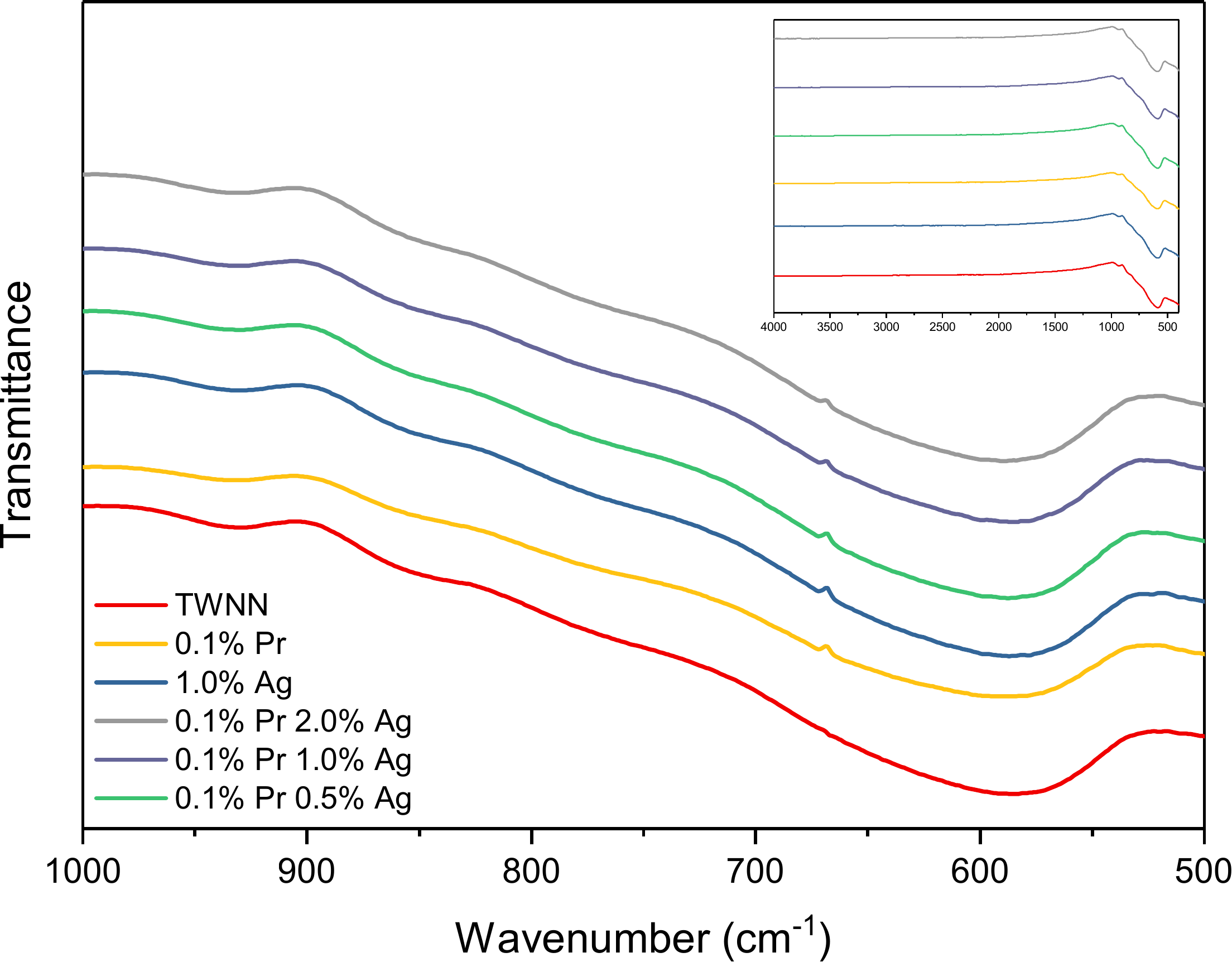}%
\caption{ATR-FTIR spectra from 1000 to 500 $cm^{-1}$ and inset from 4000 to 400 $cm^{-1}$ of all TWNN synthesized glasses.}
\label{fig:FTIR}
\end{figure}

The infrared transmission spectrum of TWNN glasses is shown in Figure~\ref{fig:FTIR}.\ Usually, glasses containing $TeO_{2}$ are composed basically of $TeO_{4}$ trigonal bipyramid (tpb) and $TeO_{3}$ trigonal pyramid (tp). According to the  crystalline tellurites, the addition of network modifiers is expected to transform part of the $TeO_{4}$ groups into $TeO_{3}$. In this case, the majority composition of the vitreous system is TeO$_{2}$, followed by the network modifier $WO_{3}$, which are the structures identified\cite{laksh2016} in Figure~\ref{fig:FTIR}. For all glasses, we identified a band centered at 595 $cm^{-1}$ which is attributed to the Te-O${^-}$ stretching vibrational in [$TeO_{4}$] trigonal bi-pyramid structure\cite{KAUR20161, Dhankhar2016}. Another band was located at 930 $cm^{-1}$, which correspond to the stretching vibrational of $W-O^{-}$ and W=O bonds in [$WO_{4}$] or [$WO_{6}$] units \cite{Dimitrov1984, VIJAY2017108, RADA20112024}. 

The OH band of the stretching mode of free Te-OH group and stretching mode of weak hydrogen bond around 3450 $cm^{-1}$ and 3100 $cm^{-1}$, respectively weren't identified, indicating that system glass have a favorable composition for $Pr^{3+}$ incorporation \cite{laksh2016, Kamalaker2010}.\ Moreover, the spectra presented the same nature for all samples, which suggests that the doping schemes did not interfered significantly on the glass structure.


\subsubsection{Absorption spectra}

\begin{figure}[!htb]
\centering
\includegraphics[scale=0.5]{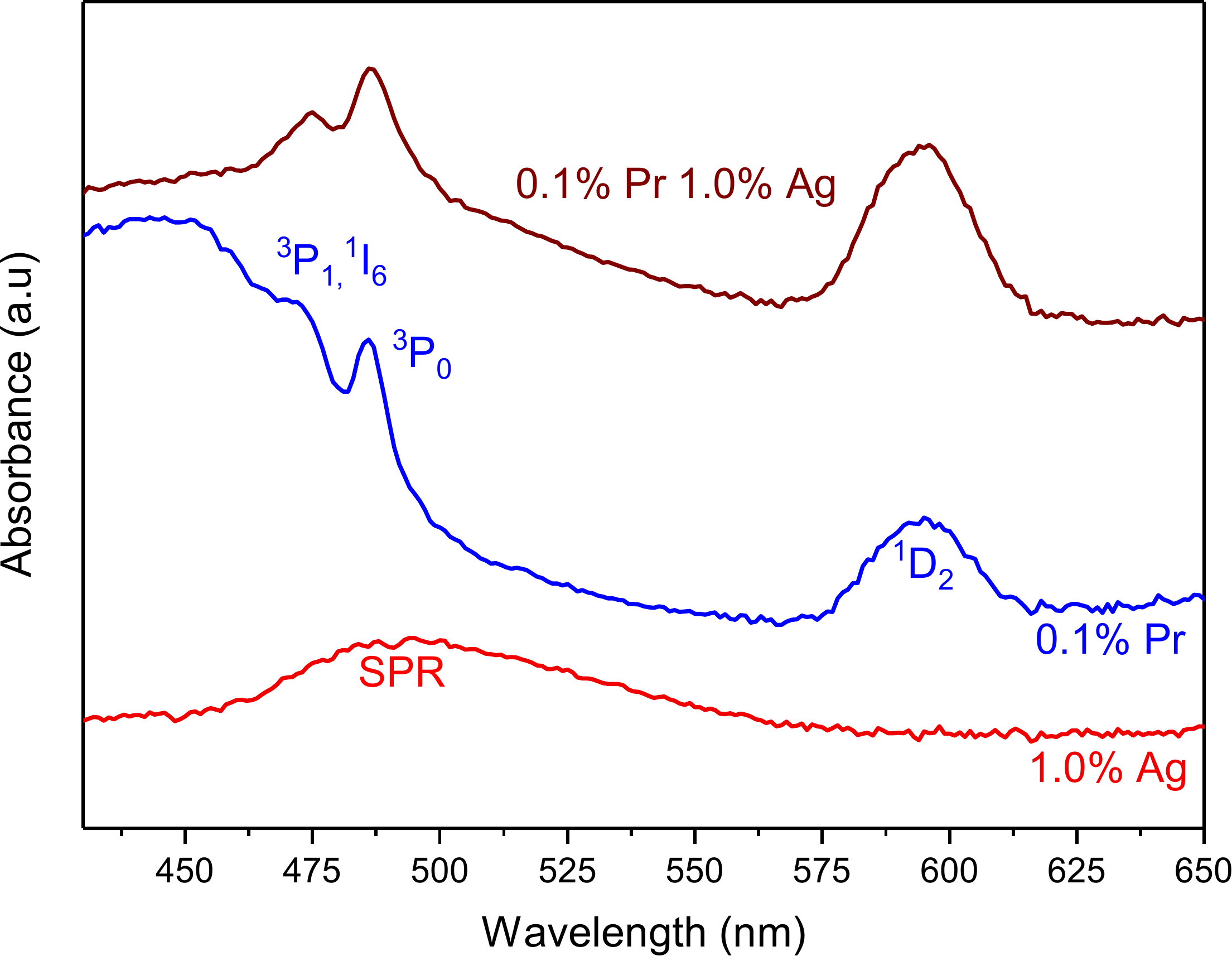}%
\caption{Absorbance spectra of TWNN samples doped with 1.0 \%mol of Ag, 0.1 \%mol and codoped with both, measured by an integrating sphere.}
\label{fig:ABS}
\end{figure}

Absorption spectra of TWNN samples containing 1.0\% of Ag, 0.1\% of Pr$^{3+}$ and codoped at these concentrations were measured using an integrating sphere and are shown in Figure \ref{fig:ABS}.

In the samples containing Pr$^{3+}$ we may see bands located at 474, 486 and 595 nm, which corresponds to the [$^{3}$P$_{1}$,$^{1}$I$_{6}$], $^{3}$P$_{0}$ and $^{1}$D$_{2}$ levels, respectively \cite{Belancon2014a, Zhou2012a}. Ag single doped sample exhibits a broad band centered at 492 nm which overlaps [$^{3}$P$_{1}$,$^{1}$I$_{6}$], $^{3}$P$_{0}$ bands in the codoped sample. 

In the literature \citeauthor{RAJESH2017607} observed the plasmon band at 492 nm in oxyfluoro tellurite glasses \cite{RAJESH2017607}, very similar to the findings of \citeauthor{CHENG2017102} (plasmon band at 510 nm in zinc tellurite glasses \cite{CHENG2017102}) and \citeauthor{Dousti2013} (plasmon band at 522 nm in sodium lead tellurite glasses\cite{Dousti2013}).\ Moreover, we conclude that the broad band observed in Ag doped samples near 500 nm in Figure~\ref{fig:ABS} are due surface plasmon resonance in Ag nanoparticles incorporated into the TWNN glass.

\begin{figure}[!htb]
\centering
\includegraphics[scale=0.5]{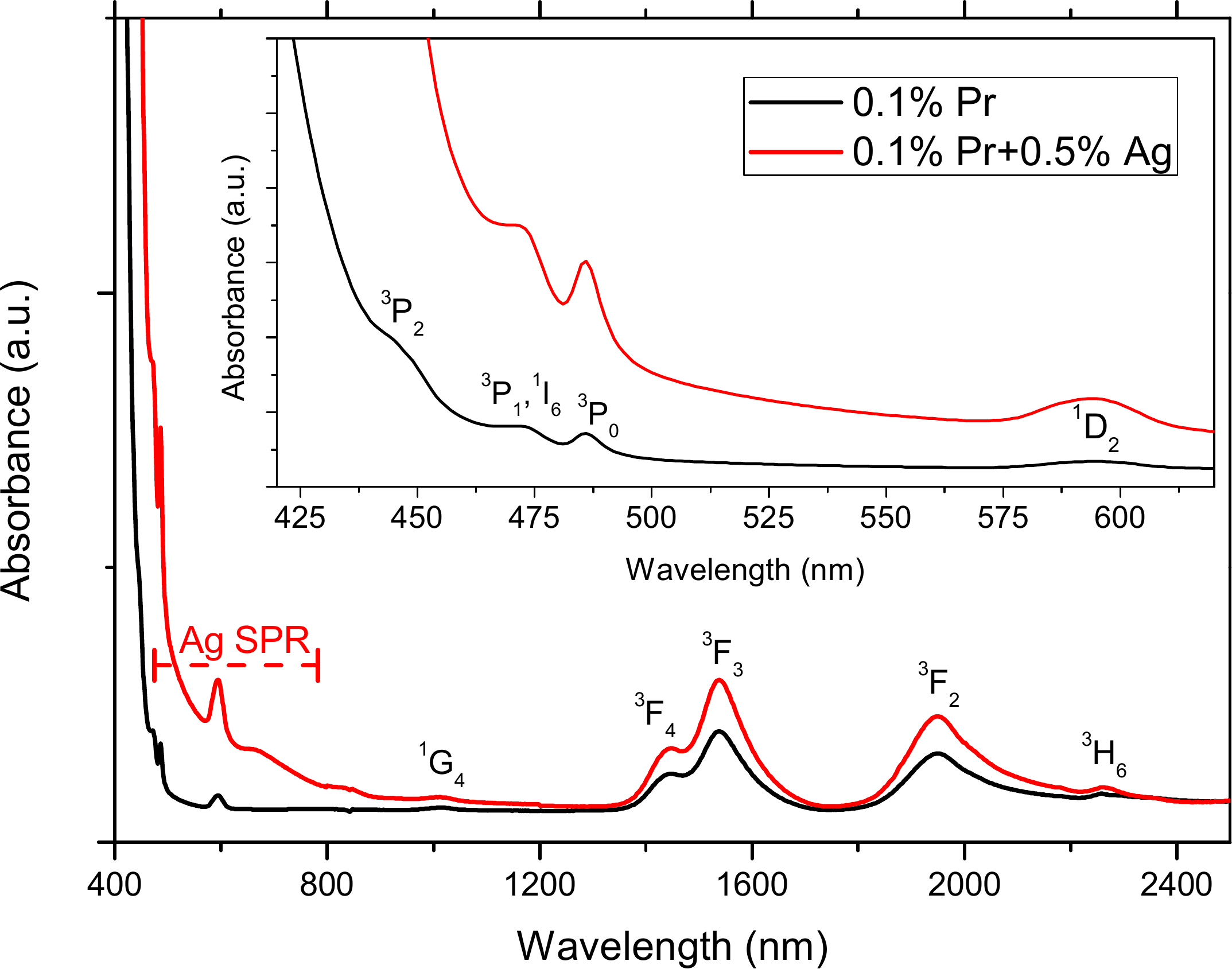}%
\caption{Absorbance spectra of TWNN samples containing 0.1\% of Pr with and without Ag, both annealed by 20 hours after quenching.}
\label{fig:ABS2}
\end{figure}

Besides this measurements performed by reflectance and scattering, samples were carefully polished and optical absorption was measured again in the range 400-2500 nm by a conventional double beam spectrophotometer in transmittance mode.\ The results for two selected samples are shown in Figure~\ref{fig:ABS2}.

As one can see, there are no detectable band similar to that one shown in Figure~\ref{fig:ABS}, however, a broad absorption band, compatible with the SPR band may be covering all the visible and the NIR spectrum.\ We highlight that it is possible that AgNPs are more concentrated in the surface of the sample\cite{Giehl2011}, or that the reflected light is more sensible to the absorption than the transmitted light. We have measured the absorption of all samples, including those submitted to different heat treatments and the SPR was at best seem as it shown in Figure~\ref{fig:ABS2}.

\subsubsection{Visible luminescence}

\begin{figure}[!htb]
\centering
\includegraphics[scale=0.5]{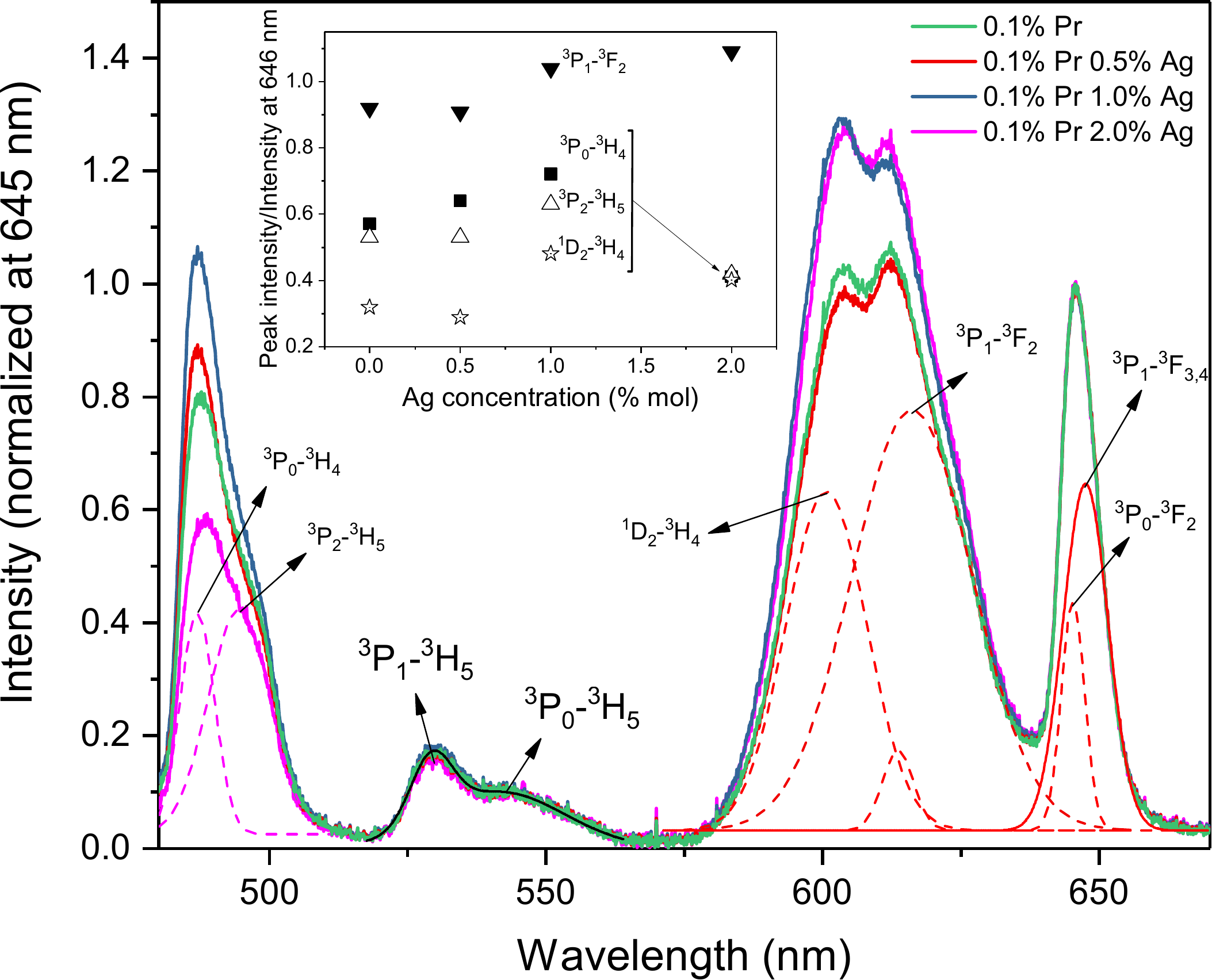}%
\caption{Emission spectra of TWNN glasses containing 0.5, 1.0 and 2.0 mol\% of Ag with 473 nm pump.}
\label{fig:PL473}
\end{figure}

Several transitions in the $Pr^{3+}$ ion may produce luminescence in the visible range of the spectrum. In Figure~\ref{fig:PL473} we show the spectra for single doped and Ag codoped samples and in the inset the peak intensities from selected transition. All samples here were heat treated by 20 hours after quenching, and all spectra are normalized by the emission intensity at 646 nm. 

Fitting the spectra with Gaussian's we could identify several bands as it is indicated in the Figure~\ref{fig:PL473}. Next, correlating their position with the gap between the levels observed in the absorption spectrum we could identify a few transitions, as well which of them are more influenced by the AgNPs. 

It is interesting to note that the bands between 525-560 nm and 640-660 nm have the same shape in all of the four samples, and this last one could not be correctly fitted by a single Gaussian, even though visually it may indicate it could. By this way, we identified two transitions in each of the regions mentioned above, which are the $^{3}$P$_{1}$-$^{3}$H$_{5}$ (529 nm) and $^{3}$P$_{0}$-$^{3}$H$_{5}$ (545 nm) for the first; $^{3}$P$_{0}$-$^{3}$F$_{2}$ (645 nm) and $^{3}$P$_{1}$-$^{3}$F$_{3,4}$ (647 nm) for the second. If this hypothesis is right, it may be the explanation of why the relative intensities in both regions remains the same, once that in both cases we have the same $^{3}$P$_{1}$ and $^{3}$P$_{0}$ upper states.

The other bands near 500 nm and 600 nm are shape sensitive to the Ag content. As a result of the analysis we have identified the transitions $^{3}$P$_{0}$-$^{3}$H$_{4}$ (487 nm) and $^{3}$P$_{2}$-$^{3}$H$_{5}$ (494 nm) as the components of the first, and the transitions $^{1}$D$_{2}$-$^{3}$H$_{4}$ (600 nm) and $^{3}$P$_{1}$-$^{3}$F$_{2}$ (615 nm) as the main components of the second. As we can see in the inset of Figure~\ref{fig:PL473}, the relative intensity of all these transitions seems to be increased with the introduction of AgNPs, however, for the highest concentration we have investigated (2.0\% Ag) the transitions do not follow the same pattern. The intensity of the transition $^{3}$P$_{1}$-$^{3}$F$_{2}$ remains at about the same level while the other three intensities have decreased. 

\begin{figure}[htb]
\centering
\includegraphics[scale=0.5]{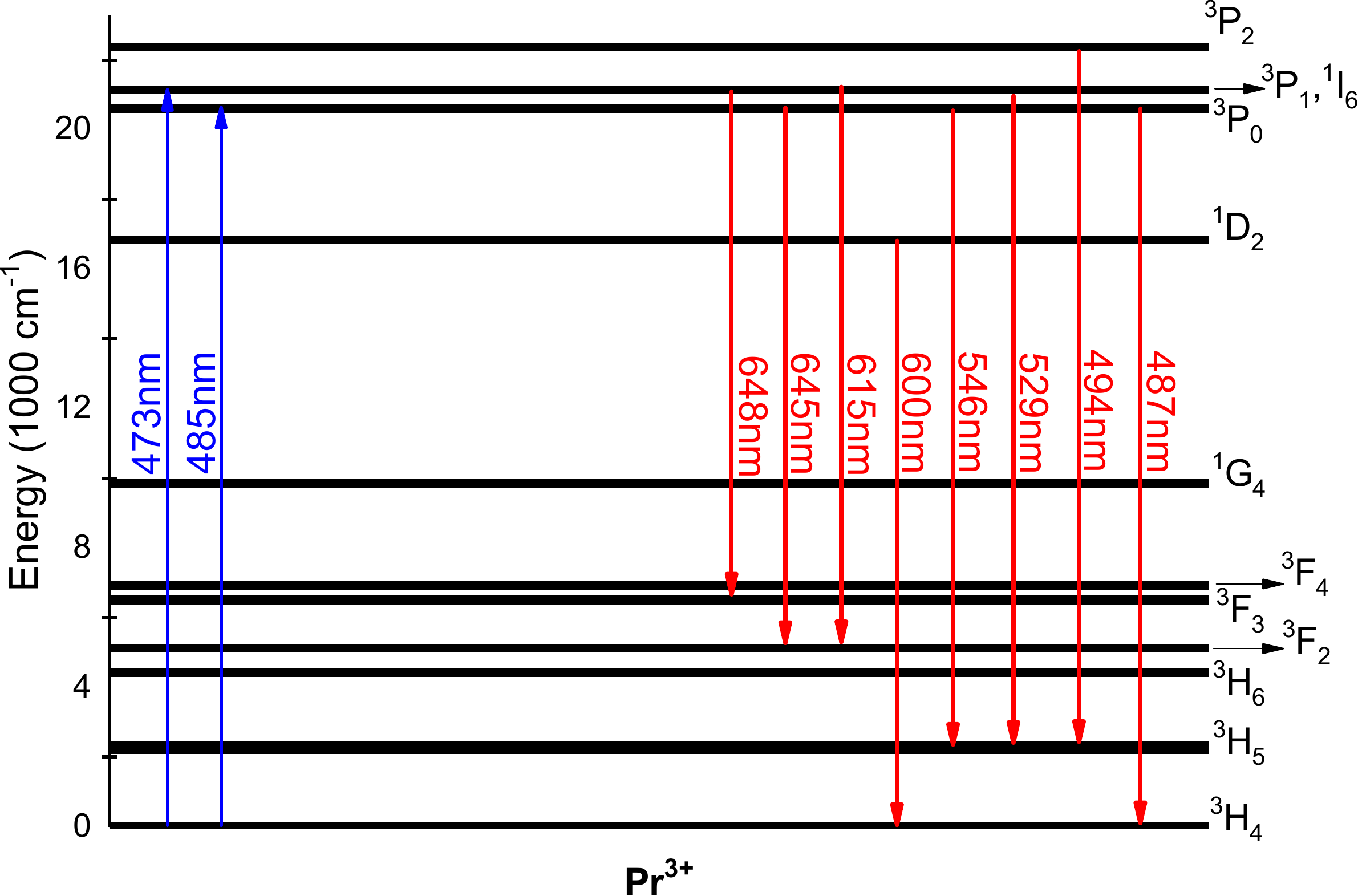}%
\caption{Energy diagram of Pr$^{3+}$.}
\label{fig:DIAGRAM}
\end{figure}

In conclusion from the spectroscopic analysis shown here, we have constructed a simplified energy diagram that is shown in Figure~\ref{fig:DIAGRAM}.

\subsubsection{Lifetimes measurements}

\begin{figure}[htb]
\centering
\includegraphics[scale=0.5]{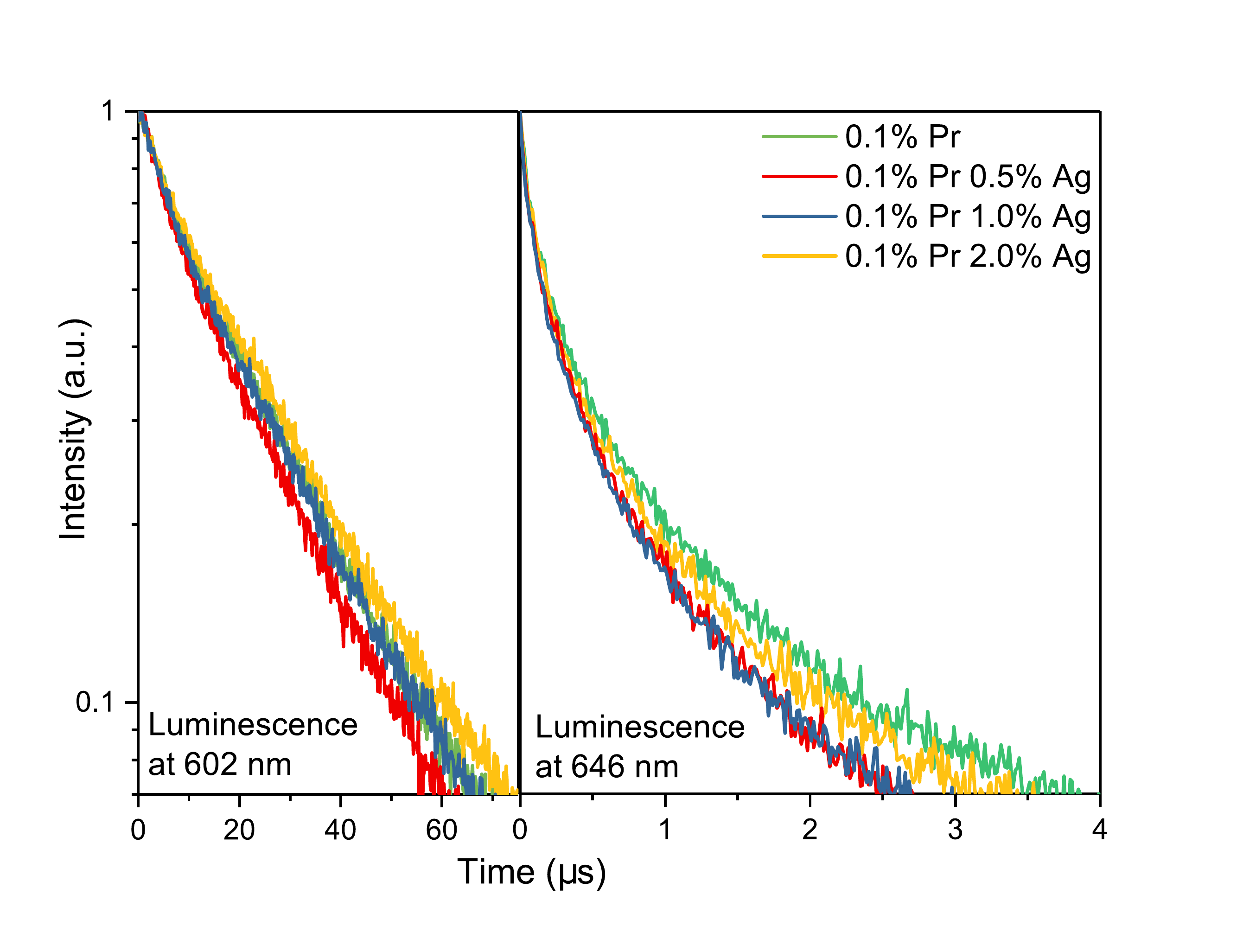}
\caption{Decays curves for $Pr^{3+}$ for samples containing 0.5, 1.0, 2.0\% and without silver with emission at 602 and 646 nm.}
\label{fig:DECAY}
\end{figure}

The decay of luminescence intensities were measured at the wavelengths of 602 nm and 646 nm and are shown in Figure~\ref{fig:DECAY}, which shows clearly that in both cases we have a non exponential decay. 

In face of this we have analyzed the curves by the integral method\cite{Reisfeld1987} and by double exponential fitting. The last one gives two lifetime constants (short $\tau_s$ and long $\tau_l$), that are expected to be similar on average to the single value obtained by the integral method\cite{Andrade2008}. In Table~\ref{tab:lifes} we may see the lifetime constant values obtained by both methods, for all intensity decays shown in Figure~\ref{fig:DECAY}.

\begin{table}[!htb]
\centering
\caption{Lifetime decay constants.}
\label{tab:lifes}
\begin{tabular}{l c c c c}
\hline
\textbf{Decays at 645nm} &$\tau_s(\mu s)$ & $\tau_l(\mu s)$ & $\tau_{int}(\mu s)$ \\
\hline

TWNN 0.1\% Pr & 0.20 & 1.83 & 1.17 \\
TWNN 0.1\% Pr 0.5\% Ag & 0.186 & 1.57 & 0.85 \\
TWNN 0.1\% Pr 1.0\% Ag & 0.19 & 1.57 & 1.13 \\
TWNN 0.1\% Pr 2.0\% Ag & 0.194 & 1.67 & 1.17 \\

\hline
\end{tabular}
\begin{tabular}{l c c c c}
\hline
\textbf{Decays at 602nm} &$\tau_s\mu s)$ & $\tau_l(\mu s)$ & $\tau_{int}(\mu s)$ \\
\hline

TWNN 0.1\% Pr & 8.12 & 28.73 & 20.23 \\
TWNN 0.1\% Pr 0.5\% Ag & 6.39 & 26.41 & 18.76 \\
TWNN 0.1\% Pr 1.0\% Ag & 7.25 & 29.36 & 20.11 \\
TWNN 0.1\% Pr 2.0\% Ag & 8.32 & 30.46 & 22.77 \\

\hline
\end{tabular}
\end{table}

\begin{figure}[htb]
\centering
\includegraphics[scale=0.5]{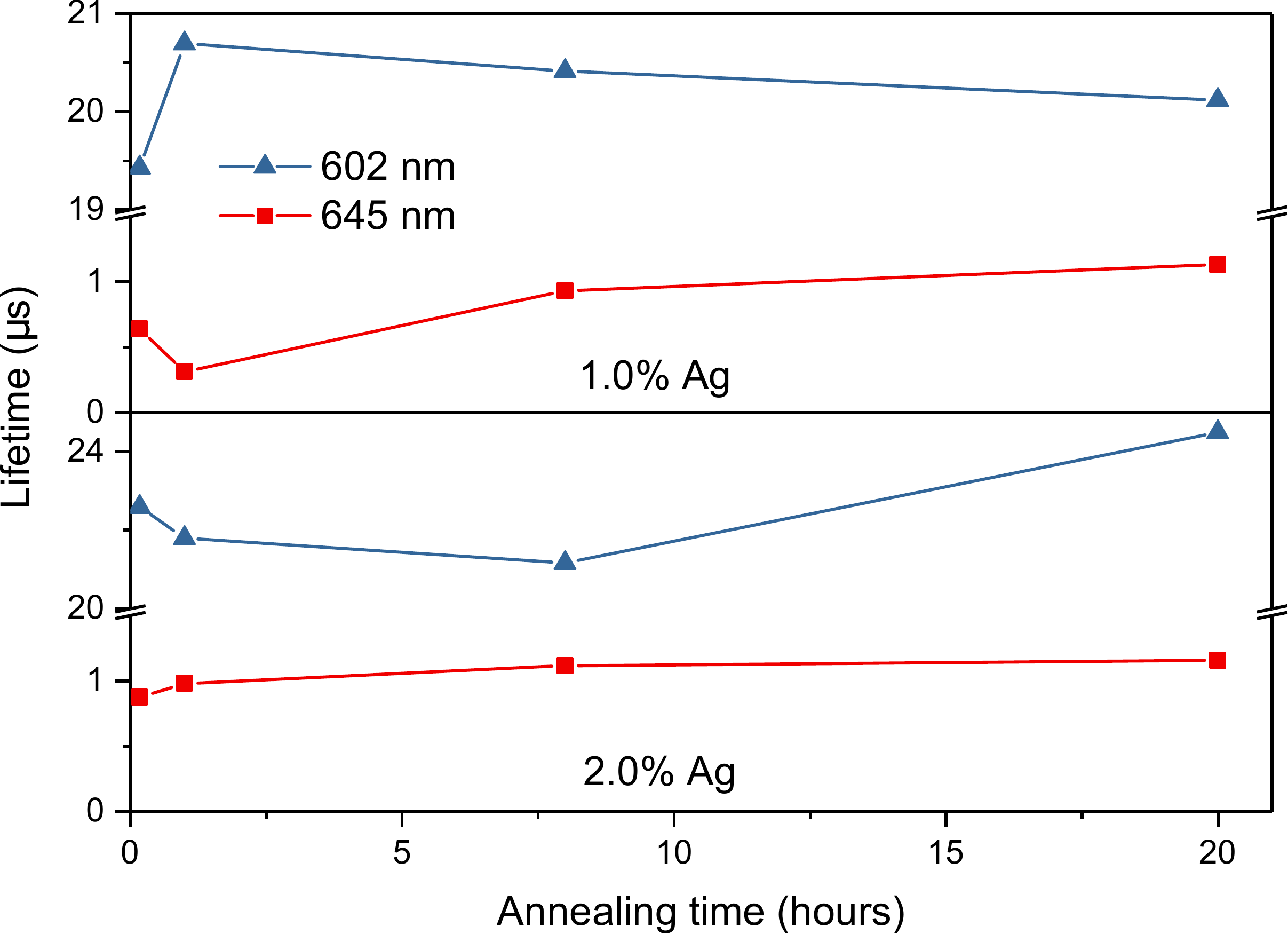}
\caption{Lifetime decay constants obtained by the integral method for samples doped with 0.1\% $Pr^{3+}$ and 1.0\% or 2.0\% of Ag.}
\label{fig:annealing}
\end{figure}

It can be observed that by both methods we are driven to the same conclusion, that higher concentration of Ag is stretching the lifetimes. The lifetime increase could be explained due to the energy transfer from Ag species to RE. This may happen because of the SPR short lifetime when comparing to the excited states of RE ions as described by \citeauthor{Malta1985}\cite{Malta1985, Malta1990}. However, at 0.5\% Ag concentration both decays are shortened. Besides that, before any attempt to find a model that could explain such results, we decided to study possible dynamics of the lifetimes induced by thermal treatment. In Figure~\ref{fig:annealing} we can see the lifetimes obtained by the integral method for two different Ag concentrations and a few different annealing times.

First of all, one may point out that these samples were produced from the same melt, which was poured in a few molds that were very near each other in the same hot plate. By this way, the only difference between these samples are how much time they were kept in the hot plate before cooling to room temperature.

So, we can conclude from Figure~\ref{fig:annealing} that lifetimes are changing during the thermal treatment, and this effect is sensible to the Ag concentration. Our analysis, which is summarized in Table~\ref{tab:lifes}, indicates that Ag is enhancing the lifetime when we look to the samples annealed by 20 hours, meanwhile one could have a different conclusion if we pick up a shorter annealing time to analyze.

Some authors have been reported effects of annealing time in the intensity and lifetimes, such as as \citeauthor{Fares2014}, that studied a tellurite glass codoped with erbium and silver \cite{Fares2014}. As the authors have discussed, once that many parameters are related to the energy transfer process between the RE and the AgNPs, such as size and distance, it can be quite difficult to identify what kind of interaction is taking place. Beyond AgNPs we may have other Ag species, which makes even more difficult to describe exactly what is going on during the annealing. 

\subsubsection{Refractive index}

\begin{figure}[htb]
\centering
\includegraphics[scale=0.5]{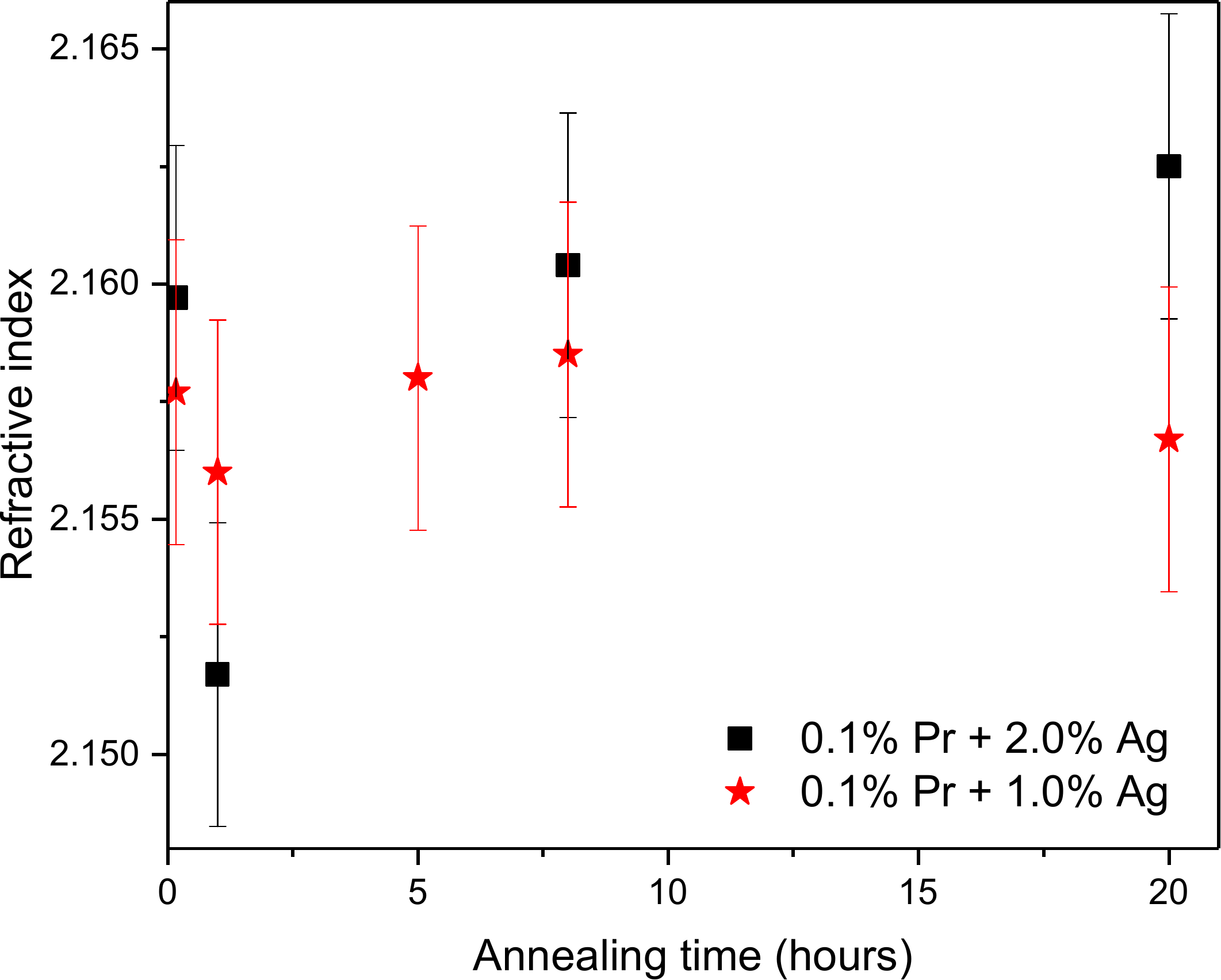}
\caption{Refractive index at 632.8 nm over annealing time}
\label{fig:index}
\end{figure}

Finally, refractive index were measured at 632.8 nm as function of the annealing time, for two different Ag concentrations, with a resolution of $\pm0.003$ (in the worst case), which was used to represent the uncertainty. The results are shown in Figure~\ref{fig:index}. The first samples were annealed by only 10 minutes, and comparing to the next ones that were annealed by 1 hour, one may see a slightly decrease in the refractive index. Next the value increases, back to the initial values after a few hours.   

Furthermore, we interpreted that Ag introduction does not produced significant structural changes into the glass. However, as the refractive index is related to the polarizability and to the number of non-bridging oxygens (NBO), the $AgNO_3$ or some derivate of it continued to be decomposed, possibly affecting the NBO's, the polarizability and by this way the refractive index\cite{Simo2012}. This is compatible with the results for the lifetimes, once that the same reaction modifying the refractive index can provide more Ag, what could contribute to increase the number of NPs or at least their sizes. 

\section{Conclusion}
\label{S:4}

TWNN glasses doped with $Pr^{3+}$ and silver nanoparticles has been synthesized and its structural, thermal and spectroscopic properties were investigated. It was observed that all the samples have a non-crystalline phase and that the addition of silver does not cause significant structural changes. In addition, the glasses showed an improvement in the thermal stability when doped with silver and absorption measurements indicates that AgNPs are present. Luminescence spectra and intensity decays have shown that NPs are interacting selectively with $Pr^{3+}$ energy levels, $\textit {i.e.}$, enhancing one transition while quenching another.\ These interactions are very sensitive to the annealing time and can be interpreted in terms of NPs growing during the process, which may change the SPR frequency and/or change the distance between $Pr^{3+}$ ions and the NPs.

Our final conclusion is that the glass phase stability and a couple of spectroscopical parameters can be enhanced due to AgNPs incorporation. However, thinking in practical applications these kind of result is not a guarantee that we may achieve improved devices, such as lasers, amplifiers or converters. Although, this remarkable phenomena observed in this research may need further investigation to evaluate the applicability and possible obstacles on the usage of those material in optical devices.


\section{Acknowledgments}
\label{S:5}
The authors would like to thank Brazilian agency CNPq (grant $480576/2013-0$) and CAPES for their financial support.












\end{document}